\documentclass[runningheads]{llncs}
\usepackage[T1]{fontenc}
\usepackage{graphicx}
\usepackage{amsmath,amssymb} 
\usepackage[subrefformat=parens]{subcaption}
\usepackage{booktabs}
\usepackage{enumerate}
\usepackage{multirow}
\usepackage{array}
\usepackage{color}

\begin{document}

\title{Increasing diversity of omni-directional images generated from single image using cGAN based on MLPMixer}
\titlerunning{Increasing diversity of omni-directional images generated from single image}
%
%
\author{Atsuya Nakata \and
Ryuto Miyazaki \and
Takao Yamanaka}
\authorrunning{A. Nakata et al.}
%

\institute{Sophia University, Tokyo, Japan \\
\email{a-nakata-7r0@eagle.sophia.ac.jp, r-miyazaki-7m7@eagle.sophia.ac.jp, takao-y@sophia.ac.jp}}
%
\maketitle              
\begin{abstract}
    This paper proposes a novel approach to generating omni-directional images from a single snapshot picture.
    The previous method has relied on the generative adversarial networks based on convolutional neural networks (CNN).
    Although this method has successfully generated omni-directional images, CNN has two drawbacks for this task.
    First, since a convolutional layer only processes a local area, it is difficult to propagate the information of an input snapshot picture embedded in the center of the omni-directional image to the edges of the image.
    Thus, the omni-directional images created by the CNN-based generator tend to have less diversity at the edges of the generated images, creating similar scene images.
    Second, the CNN-based model requires large video memory in graphics processing units due to the nature of the deep structure in CNN since shallow-layer networks only receives signals from a limited range of the receptive field.
    To solve these problems, MLPMixer-based method was proposed in this paper.
    The MLPMixer has been proposed as an alternative to the self-attention in the transformer, which captures long-range dependencies and contextual information.
    This enables to propagate information efficiently in the omni-directional image generation task.
    As a result, competitive performance has been achieved with reduced memory consumption and computational cost, in addition to increasing diversity of the generated omni-directional images.
\end{abstract}

    \section{Introduction}

    \begin{figure}[t]
        \centering
        \includegraphics[bb=0 0 808 187, width=0.9\linewidth]{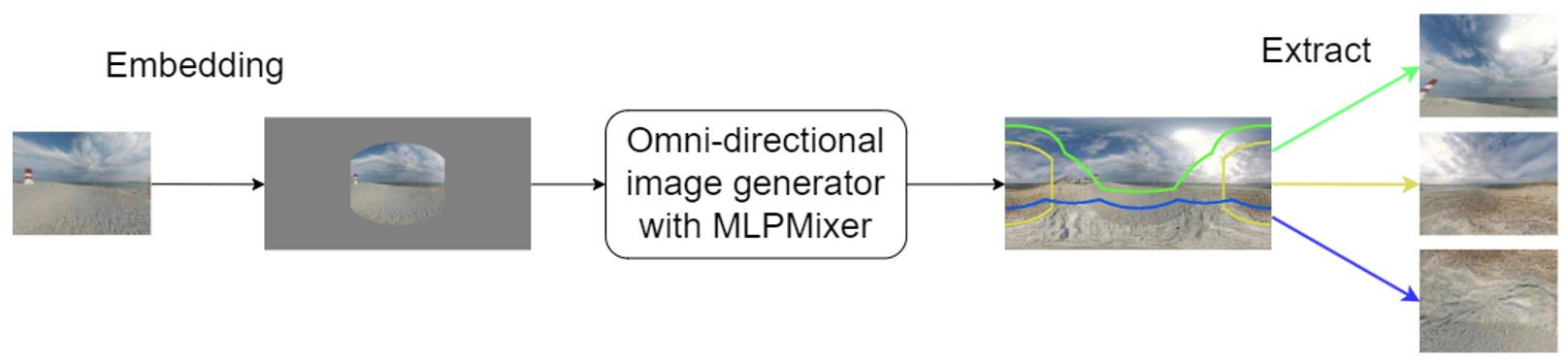}
        \caption{Outline of omni-directional image generation from single snapshot picture}
        \label{generator_image}
    \end{figure}

    An omni-directional image, or a 360-degree image, is usually captured by a camera which has a field of view covering approximately the entire sphere, and is represented in the equi-rectangular projection.
    A wide range of applications are considered such as virtual reality, driving recorders, robotics, and social networking services.
    However, the availability of these images is still limited compared with ordinary snapshot pictures, because they cannot be taken without a specialized omni-directional camera.
    To solve this problem, a method of generating the omni-directional images from a snapshot picture with the generative adversarial networks (GAN) \cite{GAN} has been proposed \cite{SphercialGAN}.
    In this method, the generator was composed of the encoder-decoder structure with convolutional layers, which was trained with the adversarial loss by inputting an embedded snapshot picture in the equi-rectangular projection as a conditional input (conditional GAN \cite{PatchGAN,cGAN}).

    Since a convolutional filter only receives the small range of local signals (small receptive field), the previous method \cite{SphercialGAN} has two drawbacks for the omni-directional iamge generation task.
    First, it has been difficult to propagate the information of the input snapshot picture embedded in the center of an omni-directioinal image as shown in Fig. \ref{generator_image} to the edges of the omni-directional image.
    Since the scene information is input to the CNN-based omni-directional image generator in the previous method \cite{SphercialGAN} as a conditional input, the edges of the omni-directional images for different input pictures for a scene might be similar scene images at the edges.
    Second, the CNN-based model requires large video memory in graphics processing units (GPU) due to the nature of the deep structure in CNN since shallow-layer networks only receives signals from a limited range of the receptive field.

    Although the self attention technique \cite{transformer} proposed for the natural language processing can represent long-distance dependencies and was successfully applied to the image recognition tasks \cite{vit}, it is prone to overfit and is not suitable for small datasets.
    An alternative architecture, MLPMixer \cite{mlpmixer}, can also represent the long distance dependencies, but can easily adjust the network size.
    With this architecture, high resolution images can be created with shallow layers by efficiently propagating the information from the center of an image to the edges.
    Therefore, this architecture was adopted in the proposed method for generating omni-directional images from a single snapshot picture.
    This property of MLPMixer also contributes to generating diversified omni-directional images conditioned on the input snapshot picture in a scene class, since the information of the input picture is efficiently propagated to the edges, generating different omni-directional images for a scene class depending on the input picture.
    The outline of the proposed method is shown in  Fig. \ref{generator_image}.
    \\\\
    The contributions of this paper include:
    \begin{enumerate}[(1)]
        \item To construct a novel architecture to generate omni-directional images from a single snapshot picture with lower memory consumption and computational cost.
        \item To design novel loss functions to appropriately train the MLPMixer-based omni-directional generator.
        \item To generate qualitatively more natural and diversified omni-directional images depending on an input snapshot picture by propagating information efficiently from the center of the image to the edges.
    \end{enumerate}

    \section{Related Works}

    \subsection{Omni-directional Image Generation}
    A method to generate omni-directional images from a snapshot picture has been proposed in \cite{SphercialGAN}.
    This method synthesizes natural landscape omni-directional images from a snapshot picture embedded in the equi-rectangular projection, by extrapolating the surrounding region using conditional GAN (cGAN) \cite{cGAN,PatchGAN}, where the conditional input is the snapshot picture.
    This cGAN is an image-to-image translation method to generate images corresponding to an input picture.
    In the previous method of generating omni-directional images, the generator has been constructed with the U-Net structure \cite{unet} composed of convolutional neural networks, which require a large memory to save a number of feature maps in the deep structure.
    This method has also proposed convolutional layers conditioned on a scene class label to train a network for all scene classes depending on the scene labels.
    In adition to this work, several researchers have also been working on the omni-directional image generation \cite{Hara_Mukuta_Harada_2021,ipoldm}.
    In \cite{Hara_Mukuta_Harada_2021}, omni-directional images have been generated using GAN and Variational AutoEncoder (VAE) by considering scene symmetry, which use a lot of convolutional layers so that it requires a large memory.
    The other work \cite{ipoldm} has proposed a method based on a latent diffusion model, which generates high quality omni-directional images from a variety of masked images.
    However, the diffusion model requires longer inference time than GAN since it requires iterative inference.

    The purpose of our paper is to improve the methods to generate omni-directional images by efficiently propagating the information of an input snapshot picture to the edges with lower computational cost and memory consumption.

    \subsection{Generative Adversarial Networks}
    A number of models have been developed to improve the quality of the generated images \cite{GANsurvey}, since the generative adversarial networks (GAN) \cite{GAN} were proposed.
    Among them, cGAN \cite{cGAN,PatchGAN} has been adopted in the previous work of generating omni-directional images \cite{SphercialGAN}, and was also used in our proposed method.
    In order to improve the stability of the training in GAN, a method to regularize the gradients of the discriminator outputs to the discriminator inputs (discriminator gradients) has been proposed \cite{wgan-gp}, by adding a term for the L2 norm of the discriminator gradients to the loss function (Gradient Penalty).
    In addition, R1 gradient penalty has been also proposed by limiting the regularization of the discriminator gradients only on the true distribution (real images) \cite{r1gp}, in contrast to R2 gradient penalty which regularizes the discriminator gradients only on the generated images.
    In our proposed method, the R1 gradient penalty was adopted to make the training stable.

    Although most of the GAN models have been based on CNN similar to the models for other tasks in the computer vision, the recent advances in the image classification task have proved that the self-attention model called transformer proposed in natural language processing (NLP) \cite{transformer} is also useful for visual image processing.
    The model called vision transformer has applied the transformer model in NLP to the image classification task by dividing the input image to patches and treating them as visual words \cite{vit}.
    The advantage of the self attention in the transformer over the convolutional layer in CNN is the ability of modeling the long-distance dependencies.
    While the convolutional  filters in CNN only receive the signals from the small local receptive field, the self attention can receive signals from every position in an image to incorporate the global information.
    Thus, the vision transformer has achieved state-of-the-art accuracy in the imagenet classification task when a large-scale training database is available \cite{vit}.


    However, one of the problems of the transformer model has been tendency to overfit \cite{vitsds} and the high computational cost, which make it difficult to be applied to the image generation in the high resolution.
    An alternative model to the self attention in the transformer has been proposed as MLPMixer \cite{mlpmixer}.
    This model first divides the input image to local patches in the similar way to the transformer, but then these signals are processed by a multi-layer perceptron (MLP) instead of the self attention.
    This enables the model to receive global information with lower computational cost than the self attention .
    These vision transformer and MLPMixer have been also applied to GAN as TransGAN \cite{transgan} and MixerGAN \cite{mixergan}.
    In our proposed method, MLPMixer was incorporated into the omni-directional image generator and discriminator.

    \section{Proposed Method}

    \begin{figure}[tb]
        \centering
        \includegraphics[bb=0 0 817 659, width=\linewidth]{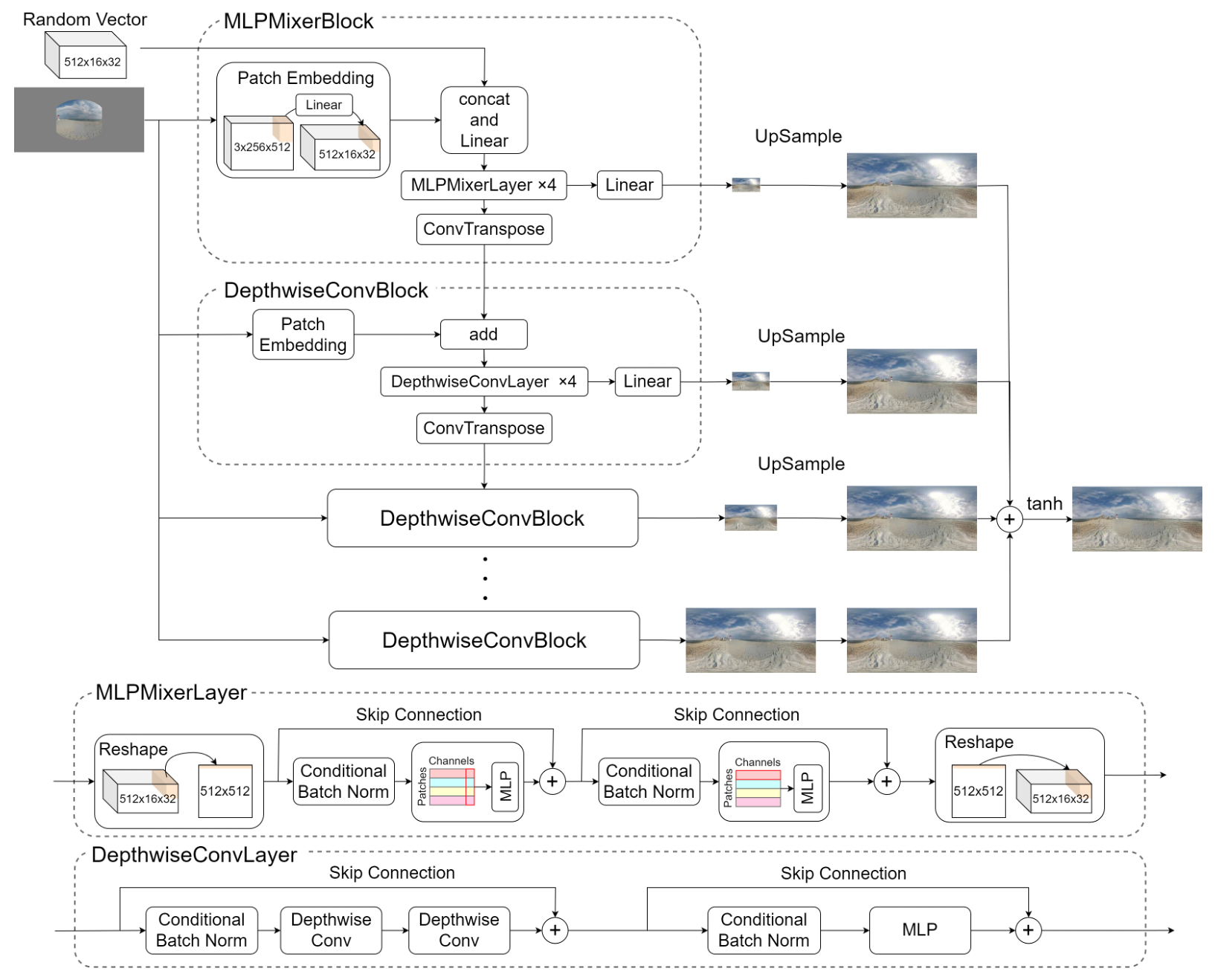}
        \caption{Proposed architecture of omni-directional image generator using MLPMixer}
        \label{model_architecture}
    \end{figure}

    \subsection{Generator}

    The architecture of the proposed model for the omni-directional image generator and discriminator is shown in Fig. \ref{model_architecture}.
    This model generates multi-scale images using the hierarchical structure, and they are integrated into an omni-directional image as the output of the generator.
    The inputs of the generator are a snapshot picture embedded in the equi-rectangular projection and a random vector sampled from a multi-modal Gaussian distribution.
    These inputs are first processed by MLPMixerBlock to produce both an output image in the low resolution and a feature map for the next block.
    Then, the feature map from the previous block and the embedded snapshot picture are processed by the DepthwiseConvBlock.
    Since the information of the snapshot picture at the center of the input image is propagated to the edges in the low resolution using MLPMixer, the following blocks do not have to use MLPMixer, and instead use the depth-wise convolutions.
    The feature maps from the MLPMixerBlock and the DepthwiseConvBlock are hierarchically processed in the following blocks to produce the omni-directional images in multi resolutions, and then are summed into an omni-directional image as the output of the generator.

    In the MLPMixerBlock, the input snapshot picture embedded in the equi-rectangular projection is processed by Patch Embedding, where the non-overlapping patches of the input image with the size of 16$\times$16 are transformed into encoded vectors using a fully connected (FC) layer (linear).
    After the channel compression in the concatenated feature map of the encoded vectors and the input random vectors, the feature map is processed with 4 MLPMixerLayers to produce the output image and a feature map for the next layer.
    The MLPMixerLayer is composed of 2 MLPs for channel-wise processing and patch-wise processing with conditional batch normalizations \cite{conditionalbatchnorm}.
    Although the original MLPMixer \cite{mlpmixer} uses the layer normalization, the conditional batch normalization conditioned on the scene class label was instead used in the proposed method to train the networks depending on the scene class label.
    After the MLPMixerLayers, the feature map is processed with PatchSplit, where the feature map is upsampled for the next block using a FC layer with reshaping the feature map.
    The output of the MLPMixerBlock is calculated by compressing the feature-map channels into 3 channels.

    The DepthwiseConvBlock is almost the same as the MLPMixerBlock except that 4 DepthwiseConvLayers without the channel compression are used instead of 4 MLPMixerLayers.
    In the $i$-th block ($i=2,...,N$) with the $N$-block structure, the conditional input of the embedded snapshot picture is divided into the non-overlapping patches with the size of $2^{N-i}\times 2^{N-i}$ to represent detailed information in the higher resolution.
    Then, the patches are transformed into encoded vectors using a FC layer, are added with the feature map from the previous block, and are forwarded to DepthwiseConvLayers.
    The DepthwiseConvLayer is also similar to MLPMixerLayer except for replacing the channel-wise MLP processing with the 2 depth-wise convolution layers, where a different convolution kernel is applied to each channel.
    In higher resolution blocks, the number of patches increases to represent the detailed information, leading to the high computational cost in the channel-wise processing with MLP.
    Therefore, this channel-wise processing with MLP is replaced by the depth-wise convolutions.
    In the depth-wise convolutions, the circular padding is used in the left and right edges to represent the continuity in the omni-directional image.

    In the experiments, the five hierarchical blocks were used to realize the omni-directional generator.
    The network structure of integrating the generated images in the multi-scale resolutions by summation was based on the structure in StyleGAN2 \cite{stylegan2}.
    The mechanism of training a network with conditional scene class labels using the conditional batch normalization was the mechanism similar to the previous work of generating omni-directional images using the conditional convolution layers \cite{SphercialGAN}.
    During the training of the proposed network, the true scene class labels from the database were used for the conditional inputs.
    For the inference, the scene class labels were estimated from the input snapshot pictures using the ResNet-based scene recognition network \cite{place365} fine-tuned to the omni-directional image database.

    \begin{figure}[tb]
        \centering
        \includegraphics[bb=0 0 806 159, width=\linewidth]{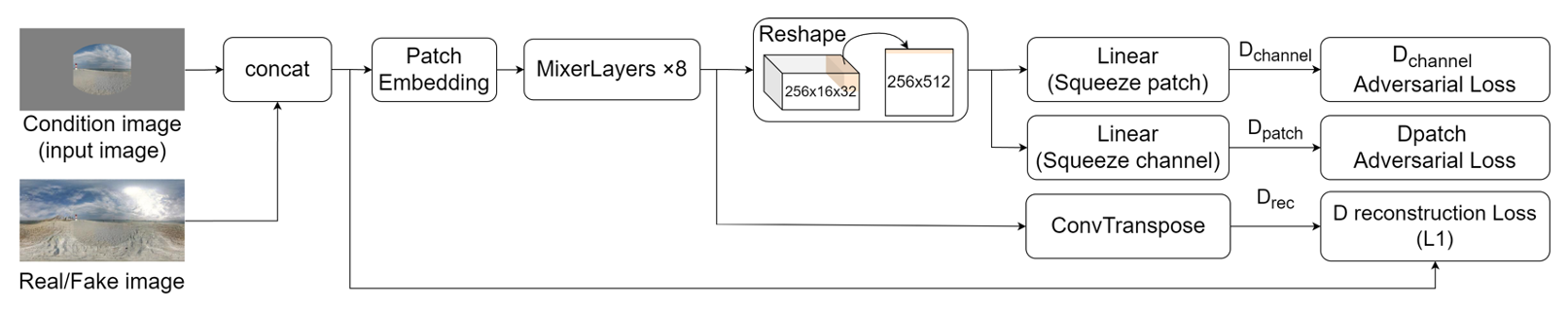}
        \caption{Structure of loss calculation in discriminator. MixerLayers are the same
                 structure as the MLPMixer Layer in the generator (Fig. 2), except that the
                 layer normalization is used instead of the conditional batch normalization}
        \label{discriminator}
    \end{figure}

    \subsection{Discriminator}
    In the generative adversarial networks, a discriminator is trained to distinguish real images and the images generated by a generator, and is used for training the generator to make the generated images as real as possible.
    The structure of the loss calculation in the discriminator is shown in Fig. \ref{discriminator}.
    The MLPMixer is also used in the discriminator by removing the global average pooling layer from the original MLPMixer networks [16] to calculate the local patch loss similar to the loss in PatchGAN \cite{PatchGAN}.
    This patch loss, $D_{patch}$ in Fig. \ref{discriminator}, is calculated by compressing the feature map from the mixer layers in MLPMixer to the single channel.
    Similar to this patch loss, the channel loss $D_{channel}$ in Fig. \ref{discriminator} is also calculated for each channel by compressing the feature map along the spatial dimension to the single patch.
    The adversarial loss in the proposed method is calculated from both the patch loss and the channel loss.
    It is noted that the layer normalizations in MLPMixer of the discriminator are not replaced by the conditional batch normalization unlike the generator.
    In addition to the adversarial loss, the reconstruction loss is also used for training the discriminator as a self-supervised regularization \cite{lwgan}.
    The output of the mixer layers in MLPMixer is upsampled using Transposed Convolution, and then is used to calculate the reconstruction loss with the real image or the generated image, as shown in Fig. \ref{discriminator} ($D_{rec}$).
    This reconstruction loss in the discriminator is based on the method in Lightweight GAN \cite{lwgan}, although the reconstruction loss is calculated only for real images in the Lightweight GAN.
    This reconstruction loss in the discriminator is also added to the loss for training the generator.
    The details of each loss in both the generator and the discriminator are explained in the next section.

    \subsection{Loss Functions}
    In the generative adversarial networks, the generator and the discriminator are trained alternately.
    The loss function for training the generator is composed of three components: the adversarial loss ($L_{adv}$), the reconstruction loss of the generator in the range where the input snapshot picture is embedded ($L_{rec}$), and the reconstruction loss of the discriminator for the regularization ($L_{dis\:rec}$).
    On the other hand, the loss function for training the discriminator is composed of two components: the adversarial loss ($L_{adv}$), and the reconstruction loss of the discriminator ($L_{dis\:rec}$).
    \begin{equation}
        \label{g_loss}
        \min_G \:L_{adv} + L_{rec} + L_{dis\:rec}
    \end{equation}
    \begin{equation}
        \label{d_loss}
        \min_D \:L_{adv} + L_{dis\:rec}
    \end{equation}
    The non-saturating loss \cite{GAN} is used for the adversarial losses in the generator.
    The loss for the discriminator is defined by the following equation.
    \\\\
    Adversarial loss for training the discriminator:
    \begin{equation}
        \label{d_adv}
        \begin{split}
            L_{adv\:m}=&\:-{E}_{x,y} \left[ \log D_m(x,y) \right] \\&- \:{E}_{x,z} \left[ \log\left( 1-D_m\left( x,G(x,z) \right) \right) \right]
        \end{split}
    \end{equation}For training the generator, the adversarial loss is usually explained by maximization of Eq \ref{d_adv}.
    Instead of maximizing Eq \ref{d_adv}, the following loss is minimized in non-saturating loss, where $D_m(x,G(x,z))$ is maximized as in the original representation but the gradient is steeper in the non-saturating loss \cite{GAN}.
    \\\\
    Adversarial loss for training the generator:
    \begin{equation}
        \label{g_adv}
        L_{adv\:m}=\:-{E}_{x,z} \left[ \log\left( D_m\left( x,G(x,z) \right) \right) \right]
    \end{equation}
    $G$ and $D$ represent the generator and the discriminator, respectively.
    $x$, $y$, and $z$ are the input snapshot picture as the conditional input, the actual omni-directional image, and a random vector.
    $m$ represents channel or patch, so that $L_{adv\:channel}$ and $L_{adv\:patch}$ are calculated from $D_{channel}$ and $D_{patch}$ in Fig. \ref{model_architecture}, respectively.
    The overall adversarial loss ($L_{adv}$) is represented by Eq. \ref{adv_loss}.

    \begin{equation}
        \label{adv_loss}
        L_{adv} = \:L_{adv\:patch} + \lambda_{ch}L_{adv\:channel}
    \end{equation}
    $\lambda_{ch}$ is a hyper-parameter to adjust the balance between the two terms.

    The reconstruction loss of the generator is calculated as the L1 norm only in the range where the input snapshot picture is embedded, given by Eq. \ref{loss_rec} using the mask $y_{mask}$ representing the calculating range and the Hadamard product ($\otimes$).
    \begin{equation}
        \label{loss_rec}
        L_{rec} = \:|\:y_{mask}\otimes (\: G(x,z)-y\:)\: |\:
    \end{equation}
    The reconstruction loss of the discriminator is also calculated as the L1 norm, but for the whole image area.
    \begin{equation}
        \label{dis_rec}
        L_{dis\:rec}=\:|\:D_{in} -  D_{rec}(D_{in})\:|
    \end{equation}
    $D_{in}$ represents the 6-channel feature map consisting of the conditional input (the embedded snapshot picture) and the actual omni-directional image $y$ or the output of the generator $G(x,z)$.
    This loss is used for the self-supervised regularization in the discriminator \cite{lwgan}.

    \subsection{Stabilization of Training}
    \label{stabilize}
    For the stabilization of the training, the differential data augmentation \cite{diffaugment} was adopted, where the data augmentation methods were applied to the inputs of the discriminator to suppress the over-fitting.
    Since the proposed method generates omni-directional images which have the property of the continuity between the left and right edges, 'roll' was applied for the shift augmentation in the horizontal direction.
    In addition to the data augmentation in the discriminator inputs, this 'roll' function was also used on the omni-directional images in the database before extracting the input snapshot pictures from the omni-directional images to increase the training data.
    In order to make the training more stable, the R1 gradient penalty \cite{r1gp} was also used during the training of the discriminator by adding the penalty term to the loss function in Eq. \ref{d_loss}.

    \section{Experimental Setup}
    The omni-directional images in 24 outdoor scenes from the SUN360 dataset \cite{sun360} were used in the experiments.
    The other outdoor-scene classes in SUN360 were excluded because there were less than 10 images in each class.
    As described in Section \ref{stabilize}, the data augmentation of the omni-directional images using 'roll' was applied before extracting the snapshot pictures.
    A snapshot picture was extracted from an omni-directional image after the augmentation, and was embedded in the equi-rectangular projection to create the databases for the training.
    After the snapshot picture was normalized in the range between -1 and 1, the surrounding region of the embedded snapshot picture was padded with 0.
    The $\lambda_{ch}$ in Eq. \ref{adv_loss} was set to 0.1 and 0.01 in the training of the generator and the discriminator, respectively.

    The CNN-based omni-directional image generator in the previous work \cite{SphercialGAN} was used as a baseline method for the comparison in the experiments.
    In this method, the generator was composed of U-Net based on CNN,
    while the discriminator also consisted of CNN based on PatchGAN \cite{PatchGAN}.
    Non-saturating adversarial loss and reconstruction loss for the generator were used for the training.

    The proposed method was trained for 200,000 iterations with the batch size of 16, while the baseline method was trained for 170,000 iterations with the batch size of 3.
    Since the performance was deteriorated after the 170,000 iterations in the baseline method, the baseline model was evaluated with the performance at the 170,000 iterations.
    The batch sizes were set to the maximum sizes with GPU of GTX1080Ti.

    \section{Results}
    After training the generator, omni-directional images were generated from the snapshot pictures extracted from the 861 test omni-directional images in the database.
    For this inference, the scene class labels were estimated using the scene recognition networks, ResNet18 \cite{place365}, fine-tuned with the images extracted from the SUN360 dataset, and then were input to the generator as the conditional information for the conditional batch normalization.
    For the evaluation, 10 snapshot pictures were extracted from a generated omni-directional image in different horizontal directions for each elevation angle of {90, 45, 0, -45, -90} degrees.
    Quantitative evaluation was conducted on Frechet Inception Distance (FID) \cite{FID}, Inception Score (IS) \cite{IS}, recognition rate of the scene label, and Learned Perceptual Image Patch Similarity(LPIPS) \cite{LPIPS}.
    FID is a metric that measures the similarity between feature distributions of real images and generated images.
    IS represents diversity and perceptual recognizability.
    LPIPS is a perceptual similarity metrics between two images, and is included for the evaluation to measure diversity of the generated images within a scene class, where the larger value means more diversified images.
    Furthermore, Multiply-ACcumulate (MAC), the memory consumption, and the inference speed were also evaluated during the inference only using the generator with the batch size 1 on CPU (Core i9-10850K) and on GPU (GeForce RTX3090).
    MAC represents the number of sum-of-product operations.
    In addition, the sample images generated using the proposed method were qualitatively compared with the generated images using the baseline method \cite{SphercialGAN}.

    \begin{table*}[t]
        \begin{minipage}[t]{\linewidth}
            \caption{Quantitative evaluation of proposed method compared with baseline.
                    Each metric was calculated for the plane images extracted from the generated
            omni-directional images at the elevation angles (90, 45, 0, -45, -90) degrees.
            }
            \vspace*{0.05cm}
            \centering
            \scalebox{0.82}{
                \begin{tabular}{lwl{2cm}wc{1.75cm}wc{1.75cm}wc{1.75cm}wc{1.75cm}wc{1.75cm}wc{1.2cm}}
                    \toprule
                    &  &  \multicolumn{5}{c}{Elevation angle [degree]} & Avg \\
                    & Metrics & 90      & 45      & 0       & -45     & -90     &      \\
                    \midrule
                    & FID($\downarrow$)  & 54.84 & 30.56 & 21.26 & 44.58 & 59.52 & 42.15 \\
                    Baseline \cite{SphercialGAN} & IS($\uparrow$)   & 3.48  & 3.07  & 3.53  & 3.78  & 3.69  & 3.51  \\
                     & Accuracy($\uparrow$)  & 24.36 & {\bf 36.41}  & {\bf 50.70}  & {\bf 45.77}  & {\bf 32.37}  & {\bf 37.92}  \\
                    & LPIPS($\uparrow$)  & 0.626 & 0.591 & 0.597 & 0.627 & 0.674 & 0.623 \\
                    \midrule
                    & FID($\downarrow$)  & {\bf 32.65}     & {\bf 20.66}    & {\bf 16.23}     & {\bf 37.24}     & {\bf 53.55}     &  {\bf 32.07}     \\
                    Proposed   & IS($\uparrow$)   &  {\bf 3.92}  &   {\bf 3.50}   &  {\bf 3.77}  &   {\bf 4.28}  &    {\bf 4.61}     &   {\bf 4.02}   \\
                     & Accuracy($\uparrow$)  & {\bf 26.64} &   32.97   &  49.08    &   43.23   &   26.82  & 35.75 \\
                    & LPIPS($\uparrow$)  & {\bf 0.648} & {\bf 0.605} & {\bf 0.622} & {\bf 0.657} & {\bf 0.724} & {\bf 0.651} \\
                    \bottomrule
                \end{tabular}
                \label{quantity_eval}}
        \end{minipage}\\
        \begin{minipage}[t]{\linewidth}
            \caption{Comparison of inference speed and model size}
            \vspace*{0.05cm}
            \centering
            \scalebox{0.85}{
            \begin{tabular}{lcccccc}
                \toprule
                & \multicolumn{2}{c}{\begin{tabular}{c}Inference speed\\\text[ms]\end{tabular}} & {Parameters} & {MAC} & {\begin{tabular}{c}Forward/backward\\pass size\end{tabular}} & {Total size} \\
                & CPU & GPU & [M] & [G] & [MB] & [MB]\\
                \midrule
                Baseline \cite{SphercialGAN} & 792.31 & 28.78 & 217.63 & 567.42 & 897.65 & 1769.73 \\
                Proposed  &  {\bf 62.19}  &  {\bf 11.03}   &  {\bf 14.40}  &  {\bf 2.14}  &  {\bf 306.27}  &  {\bf 365.42}  \\
                \bottomrule
            \end{tabular}
            \label{speed_eval}}
        \end{minipage}\\
    \end{table*}
        \begin{figure*}[t]
            \centering
            \includegraphics[bb=0 0 770 587, width=0.85\linewidth]{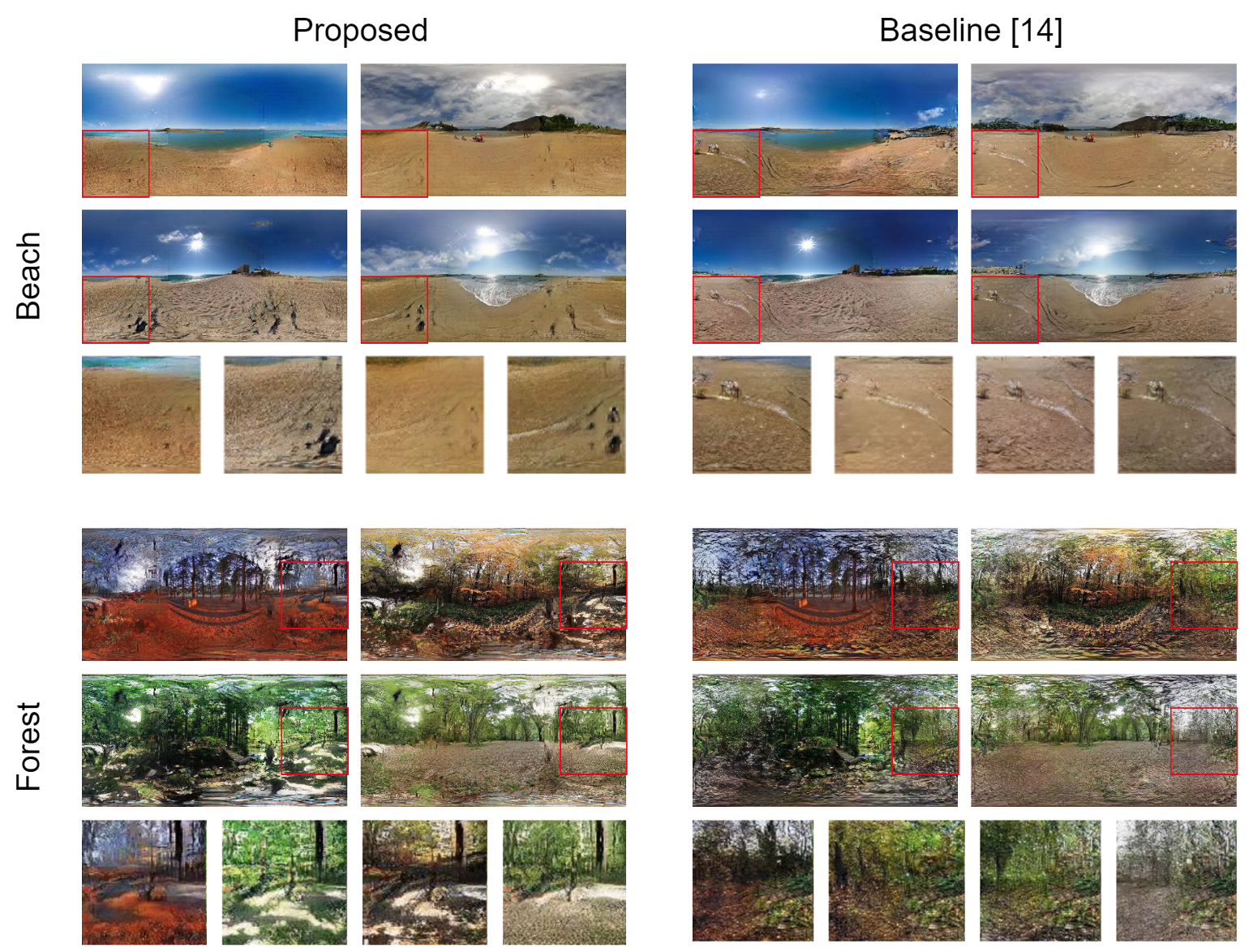}
            \caption{
                Sample images generated by proposed method and baseline method \cite{SphercialGAN}.
                For each class of 'Beach' and 'Forest', 4 sample images were generated for different input pictures
                in the same scene embedded at the center of the equirectangular projection.
                To see the diversity within the class, enlarged images of the red-frame regions are shown below the 4 sample images.
            }
            \label{quality}
        \end{figure*}
        \begin{figure*}[t]
            \centering
            \includegraphics[bb=0 0 751 128, width=0.7\linewidth]{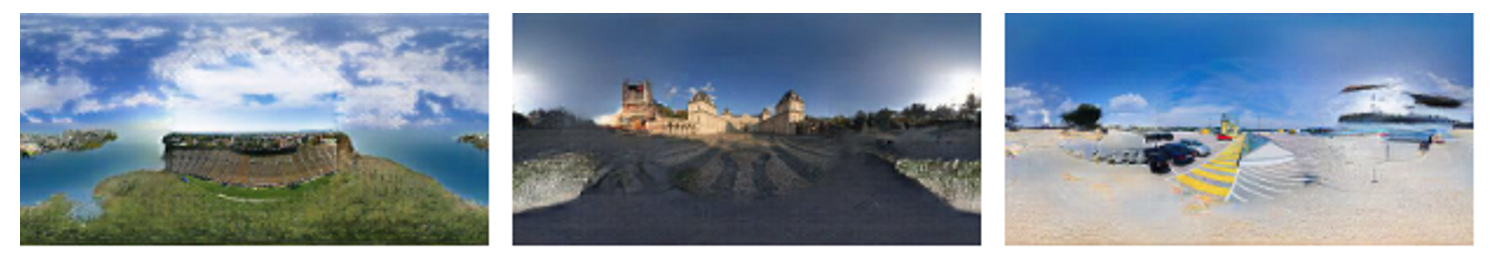}
            \caption{Limitations of proposed method.
            The information of the input pictures at the center of the images was not appropriately propagated to edges.}
            \label{low_quality}
        \end{figure*}

    \subsection{Quantitative Evaluation}
    For the quantitative evaluation, the results on FID, IS, recognition rate, and LPIPS are shown in Table \ref{quantity_eval}.
    For FID, IS, and LPIPS, the proposed method significantly outperformed the baseline method.
    Thus, the quality (FID, IS) and diversity (LPIPS) of the generated images were improved by the proposed method.
    However, the recognition rate in the proposed method was comparative or slightly lower than the baseline in many elevation angles.
    This may be due to the fact that the generated images in the baseline method tend to depend more on the scene class label because of the lack of information from an input picture near the edges than the proposed method.
    Thus, the baseline method generated similar textures at the edges of the images depending only on the scene label for various conditional input pictures.

    The results on MAC, the memory consumption, and the inference speed are shown in Table \ref{speed_eval}.
    The inference speed in the proposed method was 12.7 times faster on CPU and 2.6 times faster on GPU than the baseline method.
    The total number of parameters and MAC were significantly reduced to less than 1/15 and 1/265 of the baseline, respectively.
    The amount of memory used during the inference was greatly reduced from the baseline, which would enable inference even on devices with low GPU memories.

    \subsection{Qualitative Evaluation}
    The sample images generated by both the proposed and baseline methods \cite{SphercialGAN} are show in Fig. \ref{quality}
    for the scene classes of Beach and Forest.
    For each class, 4 sample images were generated for different input pictures in the same scene embedded at
    the center of the equiretangular projection.
    To see the diversity within the class, enlarged images of the red-frame regions of the sample images are shown below
    the 4 sample images.
    For both the scenes, similar textures were generated in the enlarged regions among 4 different input samples in the baseline method.
    On the other hand, the proposed method successfully generated different textures with various color tones and shadings.
    Thus, the proposed method increased the diversity of the generated images by propagating the information of the input
    pictures embedded at the center of the omni-directional image.

    In Fig. \ref{low_quality}, failed examples are shown for the proposed method.
    It was difficult to generate natural images for the scenes with artificial objects,
    probably due to the small number of training images.
    For these scenes, the textures for different scenes were generated around the input pictures,
    resulted in generating unnatural images for these scenes.
    A possible solution would be to increse the training data for these scenes.

    \begin{table*}[t]
        \centering
        \caption{Results of ablation studies. Metrics for $0^\circ$ of evaluation angle and Avg over elevation angles (90, 45, 0, -45, -90) were compared.}
        \vspace*{0.1cm}
        \scalebox{0.9}{
            \begin{tabular}{llwr{1.2cm}wr{1.2cm}wr{1cm}wr{1cm}wr{1.1cm}wr{1.1cm}}
                \toprule
                \multicolumn{2}{c}{\multirow{2}{*}{settings}} & \multicolumn{2}{c}{FID($\downarrow$)} & \multicolumn{2}{c}{IS($\uparrow$)} & \multicolumn{2}{c}{Accuracy[\%]($\uparrow$)} \\
                && 0°\;\; & Avg\;\; & 0°\; & Avg & 0°\;\; & Avg\;\; \\
                \midrule
                &Proposed & {\bf 16.23} & {\bf 32.07} & 3.77 & 4.02 & {\bf 49.08} & {\bf 35.75} \\
                \midrule
                (1)&\begin{tabular}{c}BN instead of Conditional BN\end{tabular} & 73.84 & 91.43 & 3.64 & 3.83 & 30.47 & 19.74 \\
                (2)&w/o $L_{dis rec}$ & 50.14 & 68.97 & {\bf 3.93} & 4.02 & 33.72 & 25.71 \\
                (3)&w/o Channel Loss & 77.13 & 98.84 & 3.87 & 4.05 & 37.15 & 26.69 \\
                (4)&w/o Multiple Inputs & 82.60 & 101.47 & 3.56 & 3.81 & 37.28 & 24.96 \\
                (5)&w/o MLPMixerLayer & 204.11 & 294.94 & 2.76 & 2.06 & 8.73 & 5.39 \\
                (6)&\begin{tabular}{c}Transformer instead of MLPMixer\end{tabular}& 19.46 & 36.29 & 3.76 & 4.06 & 44.44 & 32.59 \\
                \bottomrule
            \end{tabular}
            \label{setting_changed_eval}}
    \end{table*}

    \subsection{Ablation Study}
    Since the proposed method introduces several components to construct the model, an ablation study was conducted to investigate the influence of each component on the generated image quality.
    The hyper-parameters such as batch size were set to the same value as the proposed method without the optimization to each settings in the ablation study.
    The following settings were tested.

    \begin{enumerate}[(1)]
        \item Batch Norm (BN) instead of Conditional Batch Norm: The conditional batch normalization was replaced by the batch normalization.
        \item w/o $L_{dis rec}$: The networks were trained without the reconstruction loss for the discriminator, introduced to the proposed model for self-supervised regularization.
        \item w/o Channel Loss: The networks were trained using adversarial loss only with patch loss, but without the channel loss.
        \item w/o Multiple Inputs: The input snapshot picture was only input to the first MLPMixer block
        \item w/o MLPMixerLayer: The first block in the proposed method was replaced by DepthwiseConvLayer from MLPMixerLayer, resulted in the model without MLPMixer.
        \item Transformer instead of MLPMixer: The MLPMixer was replaced by the Transformer with positional encoding and conditional batch normalization.
    \end{enumerate}
    The results in the ablation study are shown in Table \ref{setting_changed_eval}.
    It can be seen from the results that FID and accuracy (recognition rate) were best in the proposed method, and significantly decreased without any components, although IS in some settings was comparable with the proposed method.
    This means that all the components are indispensable to achieve the performance as high as the proposed method.
    The high IS with low accuracy in the settings (2) and (3) means that a lot of omni-directional images with wrong scene classes were generated
with good image quality (high IS), so that this would indicate that the model was sensitive to the miss-classification of the scene recognition of the input snapshot pictures in the settings (2) and (3).
    Thus, the reconstruction loss for the discriminator and the channel loss would be beneficial to achieve robustness against the miss-classification.
    Among the settings, the performance was most significantly decreased in the settings (5) without MLPMixerLayer.
    This means that the propagation of the information
    from the center to the edges is the key property to obtain the high quality omni-directional images.
    The performance was also lower with Transformer (6) than with MLPMixerLayer (Proposed).
    This may be due to the fact that self-attention tends to overfit and is not suitable for small data sets.

    \section{Conclusions}
    In this paper, a novel architecture for generating omni-directional images from a snapshot picture was proposed.
    In order to propagate the information efficiently from the center to the edges, MLPMixer was adopted with the depthwise convolutions in the hierarchical structure.
    By introducing regularization terms in the loss function to make the training stable, the proposed model was successfully trained, and generated high quality and diverse omni-directional images with lower memory consumption and computational costs than the CNN-based model.
    It was confirmed that the generated omni-directional images were quantitatively and qualitatively competitive or even better than the CNN-based model.

    The proposed model would be useful not only for the omni-directional image generation, but also for out-painting tasks in general.
    One of the problems in the proposed model was the training speed due to the gradient penalty, which was calculated separately from the gradients in back-propagation.
    This would be improved by calculating the gradients both for the gradient penalty and the back-propagation at the same time, or by introducing other regularization methods to make the training stable, such as spectral normalization \cite{spectralnorm}.
    Furthermore, large-scale datasets of omni-directional images should be constructed to improve the deep-learning models in the omni-directional image processing.

    This work was supported by JSPS KAKENHI Grant Number JP21K11943.

    \bibliographystyle{splncs04}
    \bibliography{reference}

%
%
%
%
%
%
%
%
\end{document}